\documentclass[USenglish]{article}	

\usepackage[utf8]{inputenc}				
\usepackage[big,online]{dgruyter}	
\usepackage{lmodern} 
\usepackage{microtype}
\usepackage[numbers,square,sort&compress]{natbib}

\theoremstyle{dgthm}

\theoremstyle{dgdef}

\begin{document}

	\articletype{Research Article}
	\received{April 25, 2022}
  \accepted{December 5, 2022}
  \journalname{De~Gruyter~Journal}
  \journalyear{2023}
  \startpage{1}
  \aop
  \DOI{10.1515/ijb-2022-0051}

\title{Hierarchical Bayesian Bootstrap for Heterogenous Treatment Effect Estimation}
\runningtitle{Hierarchical Bayesian Bootstrap}

\author*[1]{Arman Oganisian}
\author[2]{Nandita Mitra}
\author[3]{Jason A. Roy} 
\runningauthor{A. Oganisian et al.}
\affil[1]{\protect\raggedright 
Brown University, Department of Biostatistics, Providence, RI, USA, e-mail: arman\_oganisian@brown.edu}
\affil[2]{\protect\raggedright 
University of Pennsylvania, Department of Biostatistics, Epidemiology, and Informatics, Philadelphia, PA, USA}
\affil[3]{\protect\raggedright 
Rutgers University, Department of Biostatistics and Epidemiology, Piscataway, NJ, USA}
	
	
\abstract{A major focus of causal inference is the estimation of heterogeneous average treatment effects  (HTE) - average treatment effects within strata of another variable of interest such as levels of a biomarker, education, or age strata. Inference involves estimating a stratum-specific regression and integrating it over the distribution of confounders in that stratum - which itself must be estimated. Standard practice involves estimating these stratum-specific confounder distributions independently (e.g. via the empirical distribution or Rubin's Bayesian bootstrap), which becomes problematic for sparsely populated strata with few observed confounder vectors. In this paper, we develop a nonparametric hierarchical Bayesian bootstrap (HBB) prior over the stratum-specific confounder distributions for HTE estimation. The HBB partially pools the stratum-specific distributions, thereby allowing principled borrowing of confounder information across strata when sparsity is a concern. We show that posterior inference under the HBB can yield efficiency gains over standard marginalization approaches while avoiding strong parametric assumptions about the confounder distribution. We use our approach to estimate the adverse event risk of proton versus photon chemoradiotherapy across various cancer types.}


\maketitle

\section{Introduction}
Heterogeneous treatment effects (HTEs) are causal effects within strata of some other relevant variable. These estimands are relevant in scenarios where treatment effects are believed to vary substantially in the population. In such settings, the overall estimate averaged across strata may suggest a negligible treatment effect even if there is substantial benefit/harm within a particular stratum. Modeling and estimating HTEs is important for identifying differential treatment effects in particular subgroups, which can inform targeted interventions. Such effects can be identified under rather standard causal assumptions and computed using standardization in the point-treatment setting. Within each stratum, standardization involves averaging a stratum-specific regression model adjusting for confounders and treatment over the distribution of confounders within that stratum. Fully Bayesian approaches to standardization, and causal estimation broadly, have been growing in popularity. For instance, BART regression models were used in early work by \citet{hill2011} to compute marginal effects and subsequently by \citet{zeldow2019} and \citet{henderson2018} to compute individual treatment effects and conditional average treatment effects \citet{hahn2020}. Similar approaches were developed based on Bayesian Causal Forests \citep{Caron2022,Starling2021}, an extension of BART that leverages propensity scores. Other Bayesian nonparametric (BNP) priors such as Dirichlet process (DP) mixtures and variations such as the enriched DP and dependent DP regressions have also been used to do full posterior inference on marginal treatment effects. For instance, such methods have been developed for computing effects with zero-inflated outcomes \citep{oganisian2020}, in the presence of missingness, \citep{Roy2018}, in mediation scenarios \citep{Kim2017}, for censored survival outcomes under competing risks \citep{xu2020}, and causal quantile effect computation \citep{xu2018}. Parametric Bayesian models of heterogeneity using finite mixtures were also developed in instrumental variable settings \citep{Shahn2017}.

To perform standardization, regression models must be averaged over the confounder distribution of the target population. For instance, Hill averages the BART over the empirical distribution when computing marginal effects. This is a flexible approach as it makes no modeling assumption about the distribution. However, it is unsatisfying from a Bayesian point of view since it uses a fixed plug-in estimate and uncertainty in this estimate does not flow through to the posterior of the causal effects. To overcome this issue, \citet{wang2015} and \citet{nethery2019} used the Rubin's Bayesian bootstrap (BB) \citep{rubin1981} which propagates uncertainty through to the causal effects of interest via posterior inference.

Though commonly done \citep{Roy2016,taddy2016,boatman2020}, using separate BBs for HTE estimation across strata is not ideal when some strata are sparse. For instance, in our motivating data analysis we target the marginal effect of proton versus photon chemoradiotherapy on adverse event risks. The question of interest is how the effect varies across different cancer types for which chemoradiotherapy is the standard-of-care. This is complicated as some cancer types (e.g. lung) may be rare in the sample, giving us little data on the confounder distribution within these strata. By construction, the BB places zero probability mass on confounder values unseen within this stratum - even if this is due to small samples and not due to an \textit{a priori} belief that unseen values are impossible. While plausible covariate values for lung cancer patients may have been observed for, say, brain cancer patients, stratum-specific BBs have no way of borrowing this information. There is a large literature of Bayesian models for partially pooling regression models across strata but robust nonparametric procedures for partially pooling confounder distribution models are lacking. Instead, small strata are often arbitrarily collapsed into an ``other'' category - which corresponds to a highly informative prior that the confounder distributions those small strata are equal. Another common approach is to omit estimation in those strata altogether, which does not make full use of the data.

Our main contribution is the construction of a hierarchical Bayesian bootstrap (HBB) prior for estimating stratum-specific confounder distributions in precisely such a setting. Based on the Hierarchical Dirichlet Process (HDP), our approach allows for a principled borrowing of confounder information across strata. For large strata, the HBB posterior shrinks to the stratum-specific BB. For small strata, it is shrunk more heavily towards values seen in other strata proportionally to the relative sample size of that stratum. This approach (1) maintains the flexibility of the BB (we make no parametric assumptions about the confounder distributions), (2) provides room for efficiency gains via the induced shrinkage, and (3) is fully conjugate and agnostic to the choice of outcome model. This last property makes it compatible with several of the popular outcome modeling approaches mentioned earlier.

Several notable modifications to the bootstrap have been proposed which are distinct from our work. For instance, \citet{makela2018} developed a two-stage Bayesian bootstrap for a cluster-randomized study setting. Here, clusters/strata are sampled and then individuals are sampled within a cluster. The problem of interest here is how to account for strata that exist in the population but are never sampled. This is distinct from our problem where strata are known and fixed and the issue is how to partially pool information across them. Approaches such as ``bag-of-little bootstraps'' \citep{kleiner2014, barrientos2020} have been proposed with the goal of scaling the bootstrap to large datasets. The idea is to run separate bootstraps on sub-samples, then combine in such a way as to approximate an overall bootstrap distribution. However, we are not concerned with estimating the overall data distribution, but stratum-specific distributions. Finally, several ``smoothed'' bootstraps have been developed \citep{efron1983, silverman1987, wang1995}. The view here is that the Efron's bootstrap is sampling from the empirical distribution that places uniform mass on each observed data value. This point-mass distribution is convoluted with a parametric kernel to induce smoothness. While a smoothed stratum-specific bootstrap would indeed place some mass on the unseen values, this mass is allocated via a parametric kernel, rather than informed by data in the other strata. Specification of a kernel is also a hurdle which HBB does not face. However, we can provide a probabilistic motivation for the smoothed bootstrap as an improper case of the HBB. 

In the next section, we introduce some notation and motivate the causal problem more precisely before outlining the HBB and related computation. After, we will discuss simulation studies assessing the performance of the HBB relative to dominant approaches in the causal literature under a variety of settings. We end with an analysis contrasting the risk of adverse events for proton versus photon therapies across various cancer types.

\section{Background and Motivation}
Suppose we observe outcome $Y$ for subjects assigned to treatment $A\in \{0, 1\}$ along with some confounders $L=(W, V)$ that are measured pre-treatment. These are variables which we believe to be influencing both the outcome and selection into a treatment group. In the HTE setting, this set often consists of $V$ - a discrete variable taking on values $v\in \{1, 2, \dots, K\}$ along which we wish to make causal comparisons - and variables $W$ which we would like to average over. Using potential outcomes notation \citep{Rubin1974}, one popular causal estimand is the heterogeneous, or stratum-specific, average treatment effect (HTE) $\Psi(v) = E[Y^1 - Y^0\mid V=v]$ - the average difference in outcomes had everyone in the stratum $V=v$ taken treatment 1 versus 0. This estimand is distinct from \textit{individualized}/\textit{conditional} treatment effect (CATE) estimation, which conditions on individual-level features rather than group-level to estimate $E[Y^1 - Y^0\mid L \ ]$. The connection here is that the HTE is obtained by averaging the CATE over the distribution of $W$ for each $V=v$.

While we could estimate $E[Y\mid A=a, V]$ with observed data, in general $E[Y\mid A=a, V] \neq E[Y^a\mid V]$. That is, the average outcome among subjects treated with $A=a$ in $V$ may not be the same as the average outcome had \textit{everyone} in $V$ taken treatment $A=a$. This is due to confounding: treated subjects may be a non-representative subset of the patients in stratum $V$ (e.g. systematically sicker and, therefore, more likely to have worse outcomes). Under well-known causal identification assumptions, we can estimate $\Psi(v)$ by integrating the difference in stratum-specific outcome regressions over the conditional distribution of $W$ (see Section 1 of supplement)
\begin{equation}
	\Psi(v) = \int \Big\{ E[ Y \mid A=1, V=v, w ] - E[ Y \mid A=0, V=v, w ] \Big\} dP_v(w)
\end{equation}
where $dP_v(w) = dP(w \mid V=v)$. This formula is known as standardization - a special case of the g-formula \citep{Robins1986} in the point-treatment setting. The same general approach can be used to compute an overall average treatment effect (ATE) $\Psi=E[Y^1-Y^0]$ by integrating the outcome regression over the joint $dP(l) = dP(w, v)$. The estimand $\Psi(v)$  measures strata-specific treatment effect and captures differential treatment effect by patient subgroups and so is more relevant than $\Psi$, which averages over this variability.

Suppose we observe $n$ independent subjects with data, $D=\{Y_i, A_i, W_i, V_i \}_{1:n}$. Let $S_v = \{i : V_i = v \}$ contain the indices of subjects in stratum $V=v$ and let $n_v$ denote the cardinality of $S_v$ such that $n=\sum_v n_v$. Bayesian inference typically proceeds by obtaining a posterior over $E[ Y \mid A, V, W ]$ and $P_v(w)$ which together induce a posterior over the target $\Psi(v)$. As discussed in the introduction, many BNP models exist for the former. Efficient estimation of the latter via the HBB is the chief objective of this paper, but first we review some popular alternatives. One approach is to plug in the empirical distribution $\hat P_v(w) = \frac{1}{n_v} \sum_{i \in S_v } \delta_{W_i}(w)$ - where $\delta_x(\cdot)$ denotes the degenerate distribution at $x$. For compactness we sometimes denote these as simply $P_v$ and $\delta_x$. This places uniform mass of $1/n_v$ on each confounder vectors observed in stratum $v$.

To our knowledge, \citet{wang2015} first proposed using Rubin's Bayesian bootstrap (BB) \citep{rubin1981} over this empirical approach and it has since become popular as it accounts for variability in the empirical estimate \citep{nethery2019,saarela2015,xu2018}. To summarize the BB, it models the covariate distribution as $P_v(w) = \sum_{i \in S_v} \pi_i^v \delta_{W_i}(w)$, but unlike the empirical approach the weights, $\pi^v = \{\pi_i^v\}_{i\in S_v}$,  are considered unknown parameters that completely determine $P_v$. A prior over these these weights is then a prior over $P_v$. Noting that the weight vector lives in the simplex, $\pi^v \in \{ \mathbb{R}^{n_v} : \pi_i^v >0 \ \forall i \in S_v, \ \sum_{i\in S_v} \pi_i^v = 1 \}$, the BB places an improper Dirichlet prior over this space $\pi^v \sim Dir(0_{n_v})$, where $0_{n_v}$ is the $n_v$-dimensional zero vector. This is a conjugate model with posterior $\pi^v \mid \{W_i\}_{i\in S_v} \sim Dir(1_{n_v})$, where $1_{n_v}$ is the $n_v$-dimensional vector of ones. Note that this is done for each $V=v$, \textit{separately}. This is the approach used for HTE estimation in the Bayesian causal inference literature by \citet{boatman2020}, \citet{Roy2016}, and \citet{taddy2016}.
This common approach does have several advantages. First, it retains the flexibility of the empirical distribution. Note that the posterior expectation of each $\pi_i^v$ is $1/n_v$. Second, unlike the empirical estimate, variability in this estimate flows through to the posterior of $\Psi(v)$ since the weights are not fixed at $1/n_v$. Third, it is computationally easy to sample due to conjugacy and, fourth, it is agnostic to the choice of outcome model. However, it becomes problematic for sparse strata where few values of $W$ are observed. Under the BB, $P_v$ assigns zero probability to values of $W$ that are unseen in stratum $v$. This is undesirable because there are many values that we may think are \textit{a priori} plausible. Indeed, we may observe such values in other strata. Since the BB estimates of $P_v$ are done independently, the posterior estimate of $P_v$ cannot borrow this information - yielding less stable estimates of $\Psi(v)$. In these sparse settings, it is common to collapse sparse strata into a single, pooled stratum. It is also common to simply drop subjects in sparse categories from the analysis altogether. Neither of these approaches are desirable: the former is ad-hoc and corresponds, implicitly, to a highly informative prior that the treatment effects in the sparse categories are exactly identical. The latter wastes available data. In essence, the proposed HBB retains these desirable properties of the BB while addressing the small-strata shortcomings by ``partially poolling'' the estimates of $P_v$.

\section{The Hierarchical Bayesian Bootstrap}
Let $W^v = \{W_i \}_{i \in S_v}$ denote the observed confounders in stratum $v$. We model $W^v$ as following an unknown distribution $W^v \mid P_v \sim P_v$ and propose a prior for $P_v$ that borrows information across $V$. To build intuition, first consider how this can be done parametrically. We could specify model $P_v(w) := N(w; \mu_v, 1)$ with hyper-prior $\mu_v \sim P_0(\mu_v) := N(\mu_v; \mu^*, \alpha )$ for $v=1, 2, \dots K$. Here, $P_v$ is completely determined by $\mu_v$. Centering all of the $\mu_v$ around a common prior mean $\mu^*$ allows us to borrow information across strata - which is especially helpful for the sparser ones. The parameter $\alpha$ controls the strength of information borrowing. In the improper case of $\alpha = 0$, we have $\mu_1 = \mu_2 =\dots = \mu_K = \mu^*$, which corresponds to completely pooling the data and estimating a single overall distribution. On the other hand, if $\alpha$ is large enough to be uninformative, then it is as if we estimate each of the $\mu_v$ separately - the completely unpooled case. Weakly informative values of $\alpha$ correspond to a compromise that only \textit{partially} pools the strata - with degree of pooling corresponding to degree of sparsity in the stratum. Now we consider a nonparametric extension of this idea using Dirichlet Processes. The DP is a stochastic process that generates random, discrete distributions. Due to its flexibility and conjugacy, it has become a popular prior for unknown distributions in Bayesian analysis. Suppose we place a DP prior on each $P_v$, denoted $P_v \sim DP(\alpha P_{0v})$. The realizations of $P_v$ are centered around a ``mean'' distribution of $P_{0v}$, with $\alpha>0$ controlling the dispersion of these realizations around $P_{0v}$. This is flexible because the posterior of $P_v$ under a DP is a compromise between the prior mean, $P_{0v}$, and the empirical distribution in stratum $V=v$,  $n_v^{-1} \sum_{i\in S_v} \delta_{W_i}(w)$, with relative weight controlled by $\alpha$. However, each $P_v$ is centered around its own $P_{0v}$, preventing any borrowing of information across strata. This motivates the hierarchical DP (HDP) \citep{teh2006}, which centers the $P_v$ around a \textit{common} mean distribution $P_0$ and adds a DP hyperprior on $P_0$. While the following development may seem rather involved, the actual posterior computation will be fully conjugate and efficient. Under the HDP prior, the full model for the covariates is
\begin{equation}
	\label{eq:hdp}
	\begin{split}
	W_i \mid P_v & \sim P_v \ \ \text{for} \ i \in S_v \\
	P_v \mid \alpha_v, P_0 & \sim DP(\alpha_v P_0) \ \ \text{for} \ v =1, \dots, K \\
	P_0 \mid \gamma, P_* & \sim DP(\gamma P_*)
	\end{split}
\end{equation}
The DP hyperprior on $P_0$ implies that the random $P_0$ are discrete -  allocating mass to atoms. Due to this discreteness, the distributions $P_v$ have support on the same atoms as $P_0$ but allocate mass differently across these atoms in a way that is local to $V$.  Since the DP is conjugate, the posterior of $P_v$ conditional on $P_0$ is another DP: $P_v \mid P_0, \alpha_v, W^v \sim DP( \alpha_v P_0 + \sum_{i \in S_v} \delta_{W_i})$. Similarly the marginal posterior of $P_0$ is also a DP: $P_0 \mid W  \sim DP( \gamma P_* + \sum_{i =1}^n \delta_{W_i})$. For the Hierarchical BB, we set $\gamma = 0$ in \eqref{eq:hdp} and denote this prior on $P_v$ as  $P_v \mid \alpha_v \sim HBB(\alpha_v)$. Th joint posterior under the $HBB(\alpha_v)$ is then
\begin{equation}
	\label{eq:hbbpost}
	\begin{split}
	P_v \mid P_0, \alpha_v, W^v &\sim DP( \alpha_v P_0 + \sum_{i \in S_v} \delta_{W_i}) \\
	P_0 \mid W & \sim DP(\sum_{i =1}^n \delta_{W_i})
	\end{split}
\end{equation}
With $\gamma=0$, $P_0$ are random distributions centered around the empirical distribution $P_0  \mid W \sim DP(\sum_{i =1}^n \delta_{W_i} )$ This distribution is discrete with an atom at each of the $n$ observed $W_i$. A $P_0$ can be drawn from this posterior by drawing a vector of weights $\pi_{1:n} \sim Dir(1_n)$, where $\pi_{1:n} = (\pi_1, \pi_2, \dots, \pi_n)$. This draw of $P_0$ can then be represented as $P_0 = \sum_{i=1}^n \pi_i \delta_{W_i}$. Note that this is exactly the BB. However, now we have an additional layer of uncertainty as the stratum-specific distributions must be drawn around this $P_0$: 
$ P_v \mid P_0, \alpha_v, W^v \sim DP( \alpha_v ( \sum_{i=1}^n \pi_i \delta_{W_i} )  + \sum_{i \in S_v} \delta_{W_i})$. 

\begin{figure}[h!]
    \centering
    \includegraphics[width=\linewidth]{  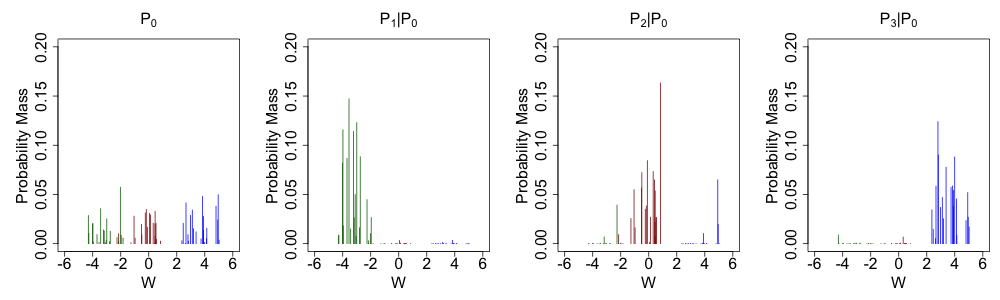}
    \caption{Draw from posterior of $P_v$ under prior $P_v \sim HBB(2)$ with simulated scalar $W_i$ for $n=90$ subjects from $V=1, 2, 3$. These 90 atoms are represented by vertical bars with colors indicating stratum of the atom. The height of the lines represent probability mass drawn from the HBB posterior. Left panel: a draw of $P_0$  - recall this is centered around the empirical distribution (i.e. line two in \eqref{eq:hbbpost} ). The next panel shows a draw from the Dirichlet Process posterior of $P_v$ conditional on this draw of $P_0$ - i.e. line one of  \eqref{eq:hbbpost}. Note that $P_1$, $P_2,$ and $P_3$ place positive mass on \textit{all} observed atoms. For instance,  independent BB estimates of $P_2$ would put place 0 mass on all atoms but the red - unlike the third panel.}
        \label{fig:simdat2}
\end{figure}

Again, conditional on a draw of $P_0$, each $P_v$ is a discrete distribution with atoms at each of the observed $n$ points in the \textit{entire} sample. Combining like terms in the summations, however, we see that atoms observed in stratum $V=v$ have a weight of $\alpha_v\pi_i + 1$ - higher than the weight on atoms unseen in stratum $V=v$, which is $\alpha_v\pi_i$. To see this, note that in expectation (over many draws of $P_v$), the posterior distribution of $W^v$ can be represented as a P\'{o}lya Urn \citep{blackwell1973}:
\begin{equation}
    \label{eq:urn}
    \begin{split}
    P_v( W = w \mid P_0, \alpha, W^v ) & = \frac{\alpha_v}{\alpha_v + n_v}( \sum_{i=1}^n \pi_i \delta_{W_i} )  + \frac{1}{\alpha_v + n_v} \sum_{i \in S_v} \delta_{W_i} \\
    							   & = \frac{1}{\alpha_v + n_v} \Big\{ \sum_{i \notin S_v} \alpha_v \pi_i \delta_{W_i} + \sum_{i \in S_v} (1+\alpha_v \pi_i ) \delta_{W_i}  \Big\}
    \end{split}
\end{equation}
Again due to the finitely many atoms, we can draw a $P_v$ from this posterior by drawing from an $n$-dimensional Dirichlet distribution with the $i^{th}$ concentration parameter being $\alpha_v \pi_i$ for $i\notin S_v$ and $1+\alpha_v\pi_i$ for $i\in S_v$. Intuitively, this can be seen as adding an additional $\alpha_v$ subjects from the marginal distribution into stratum $V$. These ``pseudo-subjects'' can take on any observed value in the marginal, even if they are unobserved in the stratum - thus, borrowing information. Similar to the parametric partial pooling example we started with in this section, here we have also partially pooled. However, rather than specifying parametric forms for $P_v$ and borrowing information in terms of their parameters, we directly partially pool the empirical distributions themselves.

As with the posterior update for $P_0$, a draw from this Dirichlet distribution yields an $n$-dimensional set of weights $\pi^v_{1:n}$ and thus a draw of $P_v$ is given by $P_v(w) = \sum_{i=1}^n \pi^v_{i} \delta_{W_i}$. We will turn to specification of hyperparameters, $\alpha_v$, after discussing computation.
\subsection{Posterior Computation via MCMC}
\label{sc:mcmc}
Here we describe posterior HTE inference under a $HBB(\alpha_v)$ prior for $P_v$ via Markov Chain Monte Carlo (MCMC). At each iterations $m=1,\dots, M$, we 
\begin{enumerate}
	\item[1.] Obtain a posterior draw of $P_0$ by drawing weights $\pi_{1:n}^{(m)} \sim Dir(1_n)$ then form $ P_0^{(m)}(w) = \sum_{i=1}^n \pi_i^{(m)} \delta_{W_i}(w) $.
	\item[2.] For each $v=1, \dots, K$,  obtain a posterior draw, $P_v^{(m)}$, conditional on $P_0^{(m)}$. We do this by drawing $\pi_{1:n}^{v(m)} \sim Dir(\eta^{(m)}_n )$, where $\eta_n^{(m)}$ is the $n$-dimensional vector with element $i$ being $\alpha_v \pi_i^{(m)}$ if $i\notin S_v$ and $(1+\alpha_v \pi_i^{(m)})$ if $i\in S_v$. Note the sum of the elements in $\eta_n^{(m)}$ is $\alpha_v + n_v$. This now forms a draw of $ P_v^{(m)}(w) = \sum_{i=1}^n \pi_i^{v(m)} \delta_{W_i}(w)$.
\end{enumerate}

Now to estimate the HTEs, suppose we also have $M$ posterior draws of the regression $E[Y \mid A, W, V]$, denoted by $\mu^{(m)}(A,W,V)$. This can be from any model. For instance, in a GLM this could be $\mu^{(m)}(A,W,V) = g^{-1} ( \beta_0^{(m)} + W' \beta_w^{(m)} + V' \beta_v^{(m)} + \beta_A^{(m)}A )$ where $g^{-1}$ is the inverse link function. This could also be a posterior draw $\mu^{(m)}(A,W,V) = f^{(m)}(A,W,V)$ where $f^{(m)}$ is the posterior draw of a sum-of-trees model under a $f \sim BART$ prior. To estimate the HTE, we include a third step
\begin{enumerate}
	\item[3.] Integrate over HBB draw of $P_v$ from Step 2, $P_v^{(m)}$.
	\begin{equation}\small
		\begin{split}
			\Psi^{(m)}(v)  &  = \int \Big\{ \mu^{(m)}(1,W_i,v) - \mu^{(m)}(0,W_i,v)  \Big\} dP^{(m)}_v(W)  =  \sum_{i=1}^n \pi_{i}^{v(m)}  \Big \{ \mu^{(m)}(1,W_i,v) - \mu^{(m)}(0,W_i,v) \Big\} 
		\end{split}
	\end{equation}
\end{enumerate}

Repeating this procedure for each of the draws yields a set of $M$ draws from the posterior of $\Psi(v)$, $\{ \Psi^{(m)}(v) \}_{1:M}$, for each stratum $v=1, \dots, K$. Note that the $W_i$ from all subjects contribute to $\Psi^{(m)}(v)$. However, values from the stratum and values outside the stratum are weighted differently according to $\pi_i^{v(m)}$.

\subsection{Some Limiting Cases and Hyperparameter Choice}
\label{sc:limitingcases}
Here we consider the limiting behavior of the HBB by analyzing \eqref{eq:urn} conditional on $P_0(w) = \sum_{i=1}^n \pi_i \delta_{W_i}$ and the choice of hyperparameter. Note that for $\alpha_v=0$, the first term in line one of \eqref{eq:urn} disappears and our estimate reduces to $P_v( W = w \mid P_0, \alpha_v, W^v ) = \frac{1}{n_v} \sum_{i \in S_v} \delta_{W_i}$. This is the empirical distribution within stratum $v$ - the posterior mean of the Bayesian Bootstrap within the stratum. It represents a \textit{completely unpooled} estimate where values of $W$ unseen in stratum $v$ have no mass. Now consider the other extreme where $\alpha_v >> n_v$. In this case \eqref{eq:urn} reduces to $P_v( w \mid P_0, \alpha_v, D ) = \sum_{i=1}^n  \pi_i \delta_{W_i}$ - the BB estimate of the overall empirical distribution (over $V$) that places expected mass $E[\pi_i] = 1/n$ on each observed value of $W$ in the entire sample. That, is we have \textit{completely pooled} all the stratum-specific distributions. The parameter $\alpha_v$ controls the posterior compromise between these extremes for a particular stratum. The idea of partial-pooling is to balance the bias-variance tradeoff, with fully pooled estimates favoring reduction in variance over potential increase in bias and fully unpooled estimates favoring a reduction of bias over potential increase in variance. Of course, partial-pooling by its nature may induce bias, especially if the confounder distributions in the sub-populations are very different. While it may be tempting to view the introduction of a user-specified parameter $\alpha_v$ as a limitation, we have just shown above that the dominant BB approach already makes a very informative prior choice of $\alpha_v = 0$ - implicitly favoring the completely unpooled scenario, even if some partial pooling to reduce variability is sensible. Introducing $\alpha_v$ makes this choice explicit and does not lock users into an implicit informative prior.
\subsubsection*{\textit{Hyperparameter guidance:}}
\begin{figure}[h!]
    \centering
    \includegraphics[width=\linewidth]{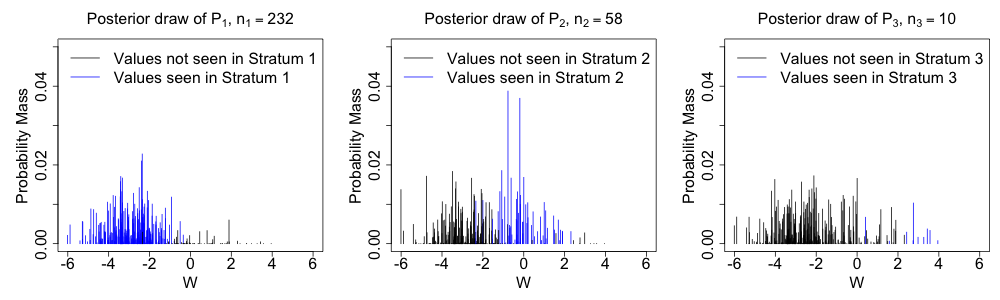}
    \caption{Draw from posterior of $P_v$ under prior $P_v \sim HBB( nM/n_v)$ with $n=300$ scalar confounders simulated for $v=1,2,3$ strata. Here we set $M=30$. Note that for stratum $V=1$ we have far greater observations than $M$ and so the draw of $P_1$ places most mass on atoms seen in this stratum. Stratum 2 has size slightly larger than $M$ and so places $\rho = 58/30 + 1 \approx 3$ times more weight on atoms seen in the stratum. Stratum 3 only has 10 subjects, and so places $\rho = 10/30 + 1 \approx 1$ equal weight on all atoms. This last case represents heaviest shrinkage.}
        \label{fig:simdat}
\end{figure}

To guide decisions about $\alpha_v$, recall that we can interpret it as adding an additional $\alpha_v$ pseudo-subjects from the marginal distribution of $W$ to the $n_v$ subjects in stratum $v$. Higher $\alpha_v$ places more weight on the pseudo-subjects - who may have values unseen in stratum $V=v$ (i.e. more shrinkage towards the marginal). The relative mass on a point seen within the stratum relative to an unseen point is approximately $ \rho = \frac{1+\alpha_v/n}{ \alpha_v/n} = \frac{n}{\alpha_v} + 1$. This is seen in \eqref{eq:urn} when substituting $\pi_i$ with its posterior expectation of $1/n$. For example, if we add $\alpha_v=n$ pseudo-subjects, then on average the atoms seen in stratum $v$ are about as likely as the atoms not seen in stratum $v$. This is fairly aggressive shrinkage. For some $M\geq 0$, one option is to set $\alpha_v = \frac{n\cdot M}{ n_v}$ which implies a relative weight of $\rho = \frac{n_v}{M} + 1$. Here, $M$ is user-specified and can be roughly interpreted as the minimum desired sample size in each stratum. This may partially be set depending on the number of confounders we are integrating over and the complexity of their joint distribution. For instance, with well-behaved, standard joint distribution (e.g. multivariate Gaussian), $M=30$ subjects within a stratum may be sufficient to estimate the distribution. On the other hands, if the covariates are complex, skewed, and multimodal we may need a larger $M$ to obtain a good nonparametric estimate such a distribution. Note that strata with size $n_v << M$ implies $\rho \approx 1$ which corresponds to heavy shrinkage. Conversely, for large strata with $n_v>>M$, $\rho$ gets larger - placing increasingly more weight on atoms within stratum $v$ only. This reduces shrinkage proportional to $n_v$. Figure \ref{fig:simdat} depicts draws from the posterior of $P_v$ under a prior $P_v \sim HBB(nM/n_v)$ with synthetic data. Note that strata that are more sparse (relative to $M$) have distributions that are more heavily shrunk towards the marginal. However, we place positive mass on all points observed in the sample.
In the supplement we outline of how full posterior updates for $\alpha_v$ can be done. While in some ways this is more satisfying, it would complicate the posterior computation with a non-conjugate update. The existing BB's popularity is due in large part to its conjugate Dirichlet updates and keeping $\alpha_v$ user-specified maintains this important property. Moreover, since the marginal posterior of $\alpha_v$ only depends on the data through $n_v$, we believe the empirical approach outlined here that sets $\alpha_v$ inversely proportional to $n_v$ is about as data-adaptive.

\subsubsection*{\textit{The smoothed bootstrap as a limiting case:}}
The smoothed bootstrap has been proposed as one way of placing mass on unseen values of $W$. In this section, we briefly show how this is a limiting case of the $HBB(\alpha_v)$ prior on a mixing distribution when $\alpha_v \rightarrow 0$. The smooth bootstrap estimate of $P_v$ is given by $\hat P_v(w) = \frac{1}{n_v} \sum_{i \in S_v}  K_h \Big( \frac{w - W_i}{h} \Big)$. Smoothness is induced by convoluting a user-specified symmetric kernel, $K_h$, with the empirical distribution and the parameter $h$ controling smoothness. For concreteness, suppose the kernel is chosen to be standard Normal $ K_h( \frac{w - W_i}{h} ) = N(  \frac{w - W_i}{h}; 0,1)$. Then this bootstrap model is a mixture of $n_v$ kernels centered around each observed $W_i$ with variance $h^2$. The mixing distribution is the empirical distribution giving weight $1/n_v$ to each mixture component. Now consider a Bayesian mixture model with \textit{unknown} mixing distribution $P_v$, written as $P(w \mid P_v ) = \int_\mathcal{W} K_h \big( \frac{w - W}{h} \big) dP_v(W) $. Here, $W$ are random with distribution $P_v$ and $w$ is a particular value. With an $HBB(\alpha)$ prior on the mixing distribution, recall that the mean of $P_v$ is given via the P\'{o}lya Urn in \eqref{eq:urn}. Plugging this urn expression in for $P_v$ yields 
$$ P(w ) = \int K_h \big( \frac{w - W}{h} \big) \Big\{\frac{\alpha}{\alpha + n_v}P_0(W) + \frac{1}{\alpha + n_v} \sum_{i \in S_v} \delta_{W_i}(W) \Big\} $$
In the improper limit as $\alpha\rightarrow 0$, the left term in the P\'{o}lya Urn goes to 0. Distributing the kernel we get $P(w ) =  \frac{1}{n_v} \sum_{i \in S_v} K_h \Big( \frac{w - W_i}{h} \Big)$. This is exactly the smoothed bootstrap estimate $\hat P_v$. Thus, we have a probabilistic motivation for the smoothed bootstrap via the HBB, formally linking our work with this previous result.

\section{Simulation Experiments}
\label{sc:sims}
Even though the HBB method is Bayesian, it is useful to examine the frequentist properties of the induced shrinkage relative to other approaches in repeated samples. Thus we conduct several simulation experiments. In all settings, we simulate 1000 datasets with $n=300$ observations from $K=4$ strata of varying sparsity. On average, the strata counts are $n_1 = 120$, $n_2=90$, $n_3=60$, $n_4 = 30$. Thus, stratum 4 is the most sparse stratum and stratum 1 is the least sparse. In each simulated data set, we simulate a vector, $W$, of 10 confounders for each subject conditional on stratum $V$. The treatment indicator $A$ itself is simulated as a function of stratum membership and confounders. We simulate a binary outcome model conditional on $V$, $W$, and $A$ from a logistic model. In the true outcome model, each stratum has a different (conditional) treatment effect, leading to true HTEs that vary across strata. The setting presented here generated the data with an outcome model containing main effects for $V$, $W$, and $L$. In Section 3.2 of the supplement we present additional settings with $W$-$A$ interactions present in each stratum's outcome model. Altogether, these settings represent challenging scenarios with several confounders and small samples that are often encountered in practice. 

For each simulated dataset, we use a correctly specified Bayesian logistic regression. This is to focus attention on the confounder distribution models. Uninformative $N(0,3)$ Gaussian priors were placed on all parameters - note this is quite wide on the log scale. We do MCMC sampling in Stan and retain 5000 posterior draws after 5000 burn-in iterations. After posterior sampling for the regression, we compute a causal risk difference, $\Psi(v) = E[Y^1 \mid V=v] -E[Y^0 \mid V=v]$ as described in Section \ref{sc:mcmc}. We integrate over four confounder distribution models. First, the empirical distribution (i.e. $\hat P_v(w) =  \sum_{i \in S_v} \frac{1}{n_v}\delta_{W_i}(w)$); Then, we integrate over the stratum-specific BB - i.e. $P_v(w) =  \sum_{i \in S_v} \pi^v_i \delta_{W_i}(w)$, where $(\pi^v_1, \pi^v_2, \dots, \pi^v_{n_v}) \sim Dir(1_{n_v})$. We also integrate over the true $P_v(w)$ via Monte Carlo, which we call the oracle. Finally, for the HBB, we set $P_v \sim HBB(nM/n_v)$ with $M=100$ in all settings. We assess the bias, variance, coverage, and precision of posterior estimates for $\Psi(1)$ and $\Psi(4)$ across simulation results in Table \ref{tab:simres}.

\begin{table}
	\caption{\label{tab:simres} Simulation results: Relative (Rel.) MSE, absolute bias, empirical variance of the posterior mean along with the width and coverage of the $95\%$ credible interval across 1,000 simulation runs. MSE is computed as average of the squared difference between posterior mean and truth across simulations. Empirical variance is computed as the variance of the 1,000 posterior mean causal effect estimates. In general, the HBB trades off bias for gains in efficiency, leading to overall reduction in MSE in stratum 4. The performance is particularly good in the Gamma mixture setting, where stratum 4 has too few observations from the tail of the Gamma distributed $P_4(W)$ to estimate it reliably via BB. The HBB, however, is able to borrow tail values observed in the other strata. }
	\centering
	\begin{tabular}{ll|ccccc|ccccc}
	\hline 
						   &              & 	\multicolumn{5}{c|}{Gaussian Mixture}           & 		\multicolumn{5}{c}{Gamma Mixture} \\
						   & Model   & Rel. MSE   & Bias & Var. & Width      & Cov.     & Rel. MSE & Bias. & Var. &  Width & Cov.   \\ \hline
		  	Strat. 1 		   & Emp.    &   	1.09	       & 0.003 & 0.005 & 0.261 & 0.938    &    0.84    & 0.013 & 0.005 & 0.268 & 0.942 \\ 
						   & BB 	  &  	1.09	       & 0.003 & 0.005 & 0.264 & 0.939    &    0.85    & 0.013 & 0.005 & 0.272 & 0.947 \\ 
						   & HBB 	  &   	1	       & 0.007 & 0.005 & 0.253 & 0.941    &       1      & 0.022 & 0.006 & 0.288 & 0.934  \\ 
					           & Oracle  &   	1.06	       & 0.004 & 0.005 & 0.259 & 0.934    &    0.77    & 0.009 & 0.005 & 0.260 & 0.950  \\  \hline
		  	Strat. 4 		   & Emp. 	  &   	1.29         & 0.003 & 0.014 & 0.465 & 0.949    &    2.93    & 0.092 & 0.023 & 0.587 & 0.904  \\ 
						   & BB 	  &    1.29         & 0.003 & 0.014 & 0.484 & 0.952    &    2.93    & 0.092 & 0.023 & 0.592 & 0.907  \\ 
						   & HBB 	  &    1              & 0.018 & 0.010 & 0.440 & 0.950    &     1        & 0.002 & 0.011 & 0.405 & 0.943   \\ 
					           & Oracle  &     1.24        & 0.000 & 0.013 & 0.463 & 0.957    &    0.88    & 0.018 & 0.009 & 0.371 & 0.933   \\  \hline
	\end{tabular}
\end{table}

The first setting considers a scenario where $W$ is marginally generated from a 10-dimensional location mixture of independent Gaussians. Thus, borrowing information from different strata is expected to come at the expense of more bias. Indeed, in stratum 4 (the most sparse stratum) we see that absolute bias is about six times higher for HBB relative to BB (.018 v .003), however variation is also lower (.01 v .014) - thus BB has an MSE 1.29 times higher overall. It is also worth noting that the HBB interval is narrower relative to BB (.440 v .484) while maintaining close to nominal coverage. In stratum 1, HBB and BB perform roughly similarly. Since this is the most populous stratum, the HBB shrinks less aggressively and produces similar estimates to the BB.

In the second setting, we consider a more complicated scenario where $W$ is generated from a 10-dimensional location mixture of Gamma distributions. Each stratum has a different mean and, importantly, skewness. This scenario is designed to assess the tail-behavior of the HBB when covariate distributions are highly skewed (e.g. income, age, etc.). As shown in Table \ref{tab:simres}, the HBB performs especially well in this complicated scenario. In stratum 4, the MSE, bias, and variance are lower than the BB. Intervals are narrower and coverage is closer to the nominal rate (94.3\%). The small sample size in stratum 4 leads to too few covariate observations from the tail of the skewed Gamma to have a reliable nonparametric estimate of $P_4(W)$. This leads to poor BB estimates with an MSE 2.93 larger than the HBB. Moreover, since it does not observe values from the full range of $W$, the BB underestimates uncertainty - yielding intervals that are too narrow and undercover. On the other hand, the HBB is able to borrow information from tail realizations observations in other strata - leading to a better estimate of $P_4(W)$. 

At a higher level, these simulation experiments demonstrate that choice of confounder distribution model may impact the causal effect estimates in small strata when confounders follow complex distributions. This supports the idea that the confounder distribution model should be thought about carefully. It also suggests that dominant default choices such as the BB and the empirical estimate, though often suitable, are not uniformly ideal. In the supplement, we provide additional simulations that show all methods perform similarly in small strata when confounder distributions are well-behaved and homogenous across strata.

\section{Adverse Event Risk of Proton versus Photon Therapy}
In this section we conduct posterior inference for casual contrasts of proton versus photon therapy among patients being treated for various locally-advanced cancers. For the cancers under consideration, standard-of-care therapy is a combination of chemotherapy and radiation - known as concurrent chemoradiotherapy (CRT). However, many modalities of radiation exist. The most common modality used in CRT has been photon radiation. In recent year, proton radiation therapy has become a more accessible alternative to patients as barriers to access have eased and health systems have adopted the necessary technology. The idea of proton therapy is to deliver radiation in a more targeted way to the cancer site, while being less damaging to healthy tissue relative to photon. Observational data were collected from $n=1468$ adult patients diagnosed with non-metastatic cancer and treated with CRT at the University of Pennsylvania Health Systems from 2011-2016.

Our data includes assigned treatment to CRT with either proton or photon radiation, several confounders measured at the time of treatment initiation, as well as the count of adverse events for a follow-up period of 90 days after treatment initiation. All patients in the sample had complete follow-up for at least 90 days. Previous research on this data \citep{baumann2020} has focused on the comparative risk of adverse events for patients on proton versus photon radiation. One hypothesis is that the more targeted nature of proton therapy will lead to fewer adverse events. Importantly, the differential risk may vary across cancer types. To address these questions, we conduct two analyses. In the first, we estimate a causal incidence difference between proton and photon patients across cancer type strata using a Poisson GLM for the adverse event count. In the second, we estimate of causal odds ratio for risk of any adverse event nonparametrically using BART. In the process we illustrate how the HBB can be combined with both parametric and nonparametric models for different outcome types. It can also be used to estimate different marginal causal contrasts (incidence differences, odds ratios, risk ratios, etc).

\subsection{Parametric Model for Causal Incidence Difference}
\begin{figure}[h!]
  \centering
  \begin{minipage}[b]{0.48\textwidth}
         \includegraphics[scale=.24]{ 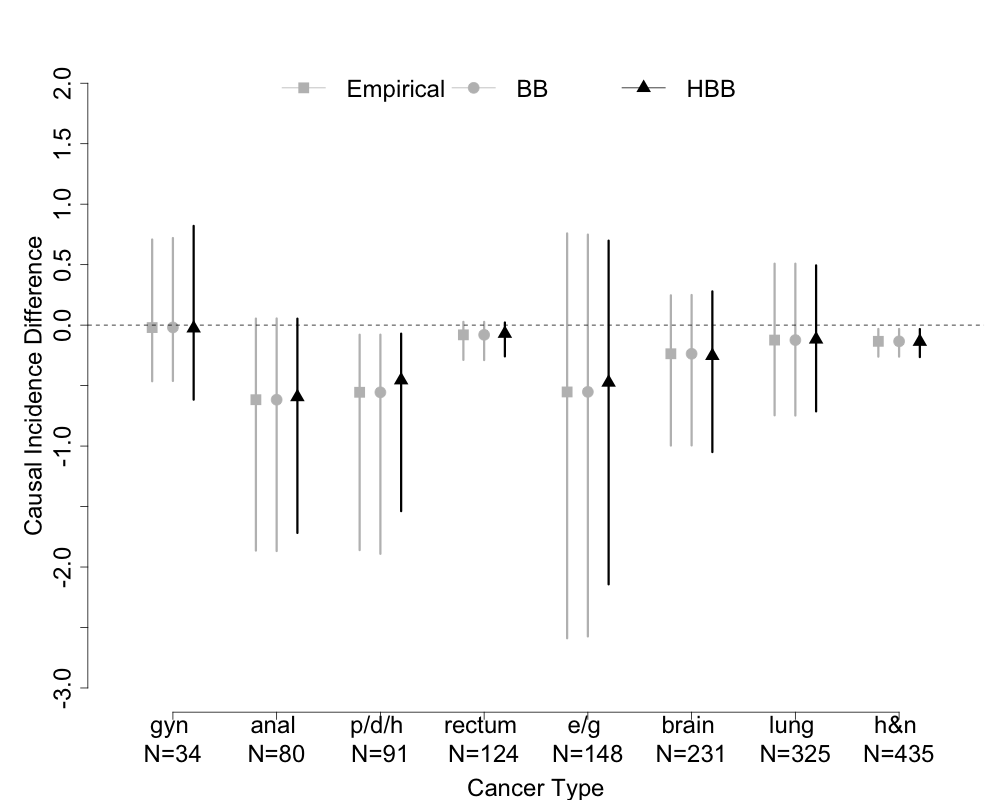}    
  \end{minipage}
  \hfill
  \begin{minipage}[b]{0.48\textwidth}
        \includegraphics[scale=.24]{ 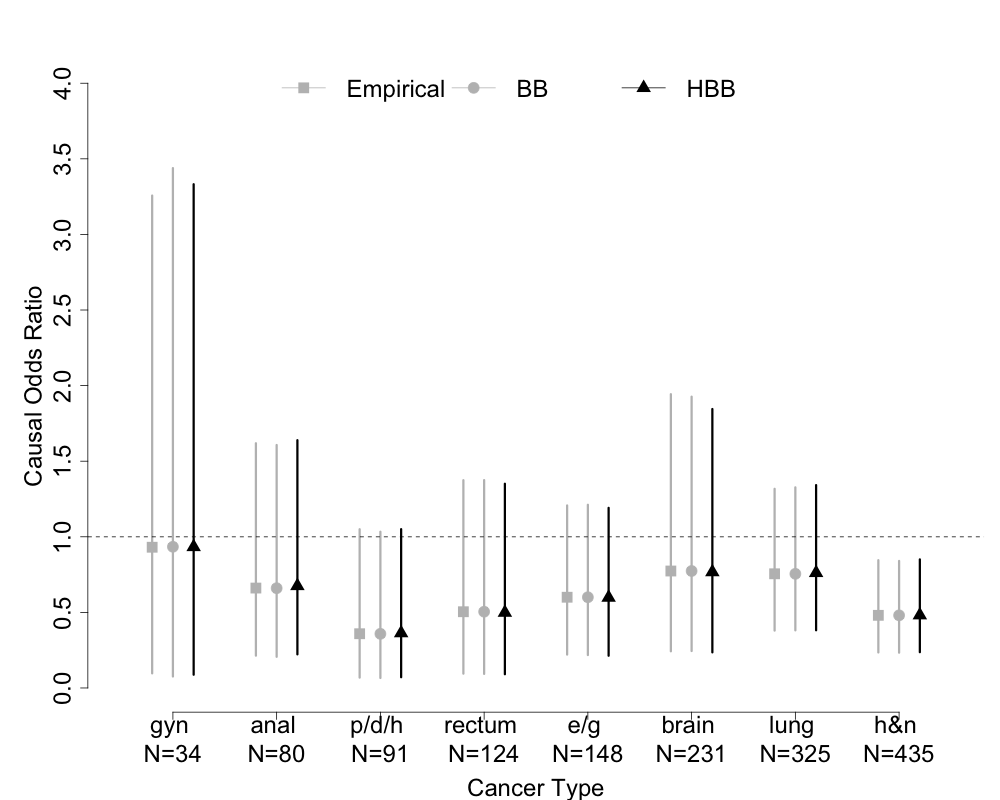} 
  \end{minipage}
  \vspace{.1in}
    \caption[test]{Posterior mean and 95\% credible interval estimates of stratum-specific causal contrasts under Poisson model (left) and BART (right). For both models, we set minimum desired sample size of $M=100$. The abbreviations are gynecological (gyn), pancreas/duodenum/hepatobiliary (p/d/h), esophagus/gastric (e/g), and head/neck (h\&n). Similar strata definitions were used in previous clinical studies \citep{baumann2020} and may be justified by anatomical closeness of affected organs.}
      \label{fig:datares}
\end{figure}

In this setting, our outcome is a count of adverse events over the 90-day follow-up, $Y\in \{0\} \cup \mathbb{Z}^{+}$. We observe data across $K=8$ cancer types (e.g., lung, head and neck, and esophagus/gastric) indicated by $V$. Let $A=1$ denote proton while $A=0$ denote photon. Finally, let $W$ be a vector of confounders including baseline age, race, sex, body-mass index (BMI), insurance plan, and charlson comorbidity index (a measure of baseline health status). We specify a conditional Poisson outcome model with the regression below. We adjust for race, sex, and insurance plan as categorical covariates. BMI, age, and charlson index are included as continuous covariates. More details on specification and prior choices are given in Section 4 of the supplement. The mean of the Poisson distribution is modeled as $E[Y_i \mid A_i, W_i, V_i =v] = \exp \{\beta^{v} + W_i' \eta^{v} + A_i\theta^{v} \}$. Though parametric, such models are common in practice. Note we allow coefficients to vary across strata. Our target of interest here is the causal incidence difference within each stratum $\Psi(v) = E[Y^1\mid V=v] - E[Y^0\mid V=v]$. A negative value indicates lower incidence of adverse events due to proton therapy relative to photon. To obtain this, we integrate the above regression over various estimators of $P_v(w)$. In the left panel of Figure \ref{fig:datares} displays results under three different estimates of $P_v$ - including the HBB (with $M=100$), BB, and the empirical distribution of $W$ in each stratum. Note that in the outcome model for gynecological cancer, we do not include sex, since this cancer occurs only in women. Even though the HBB places positive mass on male sex, since we do not integrate the outcome model along this dimension when computing $\Psi(\text{gyn})$ it has no effect. While the estimates for $\Psi(v)$ are largely similar across strata, note the HBB intervals are typically slightly shorter. Similarly, the point estimates are typically higher in these strata. This may partially reflect the trading off of increasing biased for reduced variability, as demonstrated in the simulations. However, these simulation results were averages across many runs. In any single data analysis, HBB need not produce narrower intervals.

Interpreting posterior estimates of $\Psi(v)$ in the left panel of Figure \ref{fig:datares}, we see that the proton and photon therapies' effect on adverse event incidence are largely comparable across cancer type - with posterior distributions centered either near zero or very wide around 0 (as indicated by 95\% credible intervals). Of course, these causal interpretations are subject to the validity of the required identification assumptions discussed earlier. Moreover, these inferences are conditional on the very rigid parametric assumptions. For instance, it assumes linear (on log-scale) and additive covariate effects, in addition to a poisson outcome distribution. In the next section, we consider a nonparametric estimation via BART. In the supplement, we provide results of a sensitivity analysis in which we average across HBB with various other $M$ and found the results to be largely similar.

\subsection{Nonparametric Inference for Causal Odds Ratio via BART}
Here we illustrate how the HBB can be used in conjunction with a nonparametric model for a binary outcome to obtain HTEs more robust to model misspecification. In this context let $Y\in\{0,1\}$ be a binary indicator of any adverse event over the 90-day followup period. Then, we specify a conditional Bernoulli model for $Y$ with regression $E[Y_i \mid A_i, W_i, V=v] = \Phi \big( f_v(W_i, A_i ) \big)$ with prior $f_v \sim BART$ for $v=1, \dots, K$. This is the probit specification of BART outlined in \citet{chipman2010}. Above, $\Phi$ is the standard Normal distribution function and $f_v \sim BART$ is shorthand for the sum-of-trees model $f_v(W_i, A_i) = \sum _{j=1}^J g_j^v(W_i, A_i)$ with $J$ trees, $g_j^v$. BART is characterized by a prior on the structure of each tree, $g_j^v$, consisting of terminal node parameters, splitting rules, and tree depth. Here we estimate stratum-specific models, with separate BART priors on each function. Thus, for each stratum $v$, we can get posterior draws of $f_v$ under each treatment $A=a$. In this case our target is the stratum-specific causal odds ratio $\Psi(v) = \frac{ E[Y^1\mid V=v] /(1-E[Y^1\mid V=v])  }{ E[Y^0\mid V=v] /(1-E[Y^0\mid V=v]) }$. Values of $\Psi(v)$ less than one indicate lower risk of any adverse event due to proton therapy, relative to photon. Using standardization, we can compute each expectation by integrating $ \Phi( f_v(W, a) )$ over $P_v(W)$. The right panel of Figure \ref{fig:datares} displays posterior results for $\Psi(v)$ under three different estimate of $P_v$ - including the HBB (with M=100), BB, and the empirical distribution of $W$ in each stratum. We notice that while point and interval estimates are generally similar across strata, the HBB intervals are somewhat narrower. However, according to these results, there is little posterior evidence for a reduction of adverse event risk due to proton therapy. While point estimates of the odds ratios are below one across strata, there is significant posterior uncertainty about the direction and magnitude of these effects, as indicated by the wide 95\% credible intervals mostly overlapping one. In order to better compare these results with those from the previous model, in the supplement we provide the corresponding causal odds ratios of any adverse event computed from the Poisson model. The results are largely similar across BART and Poisson, with the Poisson intervals in some strata being slightly narrower.

\section{Discussion}
The confounder distribution is a key unknown that must be estimated flexibly when making causal inferences. It is still more important in the context of HTEs where some strata may be too sparse to allow reliable nonparametric estimation. In this paper we show that straightforward application of the Bayesian bootstrap, though common, can be improved upon in these scenarios with the HBB. The proposed HBB shares covariate information across strata to achieve more stable stratum-specific causal estimates. The approach is computationally tractable, compatible with arbitrary outcome models, and makes no parametric assumptions about the distributions. As shown in the data analysis, it can be used to compute a variety of marginal causal contrasts. 

In simulations, we show that in populous strata, the HBB performs in-line with the BB and the empirical distributions. We expect this to be the case since, as shown in Equation 4, the HBB approaches the BB as $n_v$ gets large. We also show in additional simulations in the supplement that the HBB performs comparably with these other approaches in sparse strata when the true confounders distributions are the same across strata. We see a significant advantage to the HBB in sparse strata with complex confounder distributions. In these strata, we may not have enough observations to capture all the complexities. The Gamma mixture setting demonstrates one version of this scenario where we have skewed confounder distributions and too few realizations from the tail of the distribution. The fact that the BB has difficulty in capturing tails has been known for some time. In the original BB paper \citet{rubin1981} writes: ``Is it reasonable to use a model specification that effectively assumes all possible distinct values of $X$ have been observed?...consider the probability that $X>c$ where $c$ is larger than the largest observed $X$...the simple BB and bootstrap estimate such probabilities as 0 with zero variability, which is clearly inappropriate.'' He notes that because of this ``inferences about moments will be sensitive to the model misspecification of tail probabilities.'' To be precise, consider a simple setting with univariate continuous real-valued $L$ and recall that the average casual difference in stratum $V=v$ is $\Psi(v) = \int \Psi(v, w) P_v(w)dw$, where $\Psi(v, w) = E[Y\mid A=1, V=v,W=w ]-E[Y\mid A=0, V=v, W=w] $. Suppose the minimum and maximum of the observed $W$ in stratum $V=v$ are $w_{(1)}$ and $w_{(nv)}$ respectively. Then, 

$$\Psi(v)  =  \int_{-\infty}^{w_{(1)} } \Psi(v, w) P_v(w)dw + \int_{w_{(1)} }^{w_{(nv)} } \Psi(v, w) P_v(w)dw +   \int_{ w_{(nv)} }^ {\infty}  \Psi(v, w) P_v(w)dw $$

Using the Bayesian bootstrap or empirical distribution may allow decent estimation of the middle term - but the left and right terms will be estimated as $0$ since $P_v(w)$ allocates no probability mass to unobserved values. This is true even if the outcome model is correctly specified. This shows how estimates of the average causal effect can be sensitive to the tails of the confounder distribution. The HBB shines in these settings by leveraging tail values observed in other strata. This allows for better estimation of the right and left terms in the equation above and thus overcomes an important failure-point in the standard bootstrap. 

We emphasize that potential applications of the HBB go beyond estimation of stratum-specific average causal effects. For instance, another popular causal estimand is the average treatment effect on the treated (ATT). This is defined as the average difference in potential outcomes among those assigned treatment. A Standardization-type procedure can be used here as well and requires integrating a regression over the distribution of confounders among the treated, $P(W \mid A=1)$. If there are too few treated subjects to get a reliable nonparametric estimate of this distribution, it may be reasonable to borrow covariate information from untreated subjects, $P(W \mid A=0)$, by shrinking towards the marginal via the HBB. 

Computing causal effects within several strata may raise issues of multiplicity. In the frequentist framework, this arises in the context of multiple hypothesis testing and is addressed through various corrections to control a family-wise error rate within the set of tests. In the Bayesian framework, even though we are not conducting hypothesis tests, multiplicity-type issues may arise when we estimate effects across many strata and run into some strata with much larger estimates due to small sample variability. One common Bayesian solution is to use weakly informative null-centered priors on the stratum-specific effects that shrink posterior estimates towards the overall average. This penalizes extreme estimates \cite{Gelman2012} in sparse strata.

Lastly, our discussion of the connection between the HBB and the smoothed bootstrap motivates an extension to a ``smoothed HBB''. In Section \ref{sc:limitingcases}, an $HBB(0)$ prior on the mixing distribution corresponds to a smoothed bootstrap within a stratum but prevents borrowing of information. In principle setting $\alpha_v>0$ would yield a posterior that is a hierarchical DP mixture of $K_h$ - thus borrowing information across strata while modeling the distribution as a smooth mixture. If, for instance, $K_h$ is a Gaussian kernel, we speculate the strength of the shrinkage could be informed by the $L$-$2$ distance in covariate values across strata. While such distance-based shrinkage would be appealing, posterior computation would be much more involved - requiring updating the kernel parameters as well as good default choices of $K_h$. An advantage of the HBB is that we require no specification of distance metric/kernel and maintain computational ease. However, this extension would be interested to pursue in the future.
	
\begin{acknowledgement}
We would like to thank James Metz and Justin Bekelman (Department of Radiation Oncology, Perelman School of Medicine, University of Pennsylvania) for data support.
\end{acknowledgement}

\bibliographystyle{rss}
\vspace{-.3in}
\bibliography{references}
\vspace{-.3in}

\end{document}